\documentclass[12pt]{article}

\usepackage{amsfonts, amsmath, amssymb, amsthm}
\usepackage{a4}

\DeclareMathOperator{\Trace}{Trace}

\theoremstyle{definition}

\theoremstyle{remark}

\author{I.G. Korepanov}
\title{``Nonconstant cohomology'' of Hietarinta's two-color solutions to four-simplex equation}
\date{January--February 2016}

\begin{document}

\maketitle

\begin{abstract}
``Nonconstant cohomologies'' are introduced for solutions of set-theo\-ret\-i\-cal four-simplex equation (FSE). While usual cohomologies lead to solutions of constant quantum FSE, our ``nonconstant cohomologies'' lead to solutions of nonconstant quantum FSE. Computer calculations are presented showing that large spaces of such cohomologies exist for all Hietarinta's two-color linear solutions to set-theoretical FSE. After taking a partial trace of the corresponding quantum operators, combined with one additional trick, this leads to solutions of tetrahedron equation, including those with non-negative matrix elements, and not reducible to a permutation, even with cocycle multipliers.
\end{abstract}

\section{Four-simplex equation}\label{s:fse}

\subsection{Set-theoretical FSE}\label{ss:stfse}

Let there be a set~$X$, whose elements are called `colors', and a map
\[
R\colon\; X\times X\times X\times X\to X\times X\times X\times X
\]
from its fourth Cartesian degree into itself. We will need also the \emph{tenth} Cartesian degree~$X^{\times 10}$ of~$X$, and, to distinguish between the ten Cartesian factors --- copies of~$X$ --- we will call them also $X_1,\dots,X_{10}$. Let $1\le i<j<k<l\le 10$; we denote
\[
R_{ijkl}\colon\; X_i\times X_j\times X_k\times X_l \to X_i\times X_j\times X_k\times X_l
\]
the copy of~$R$ acting in the product the corresponding copies of~$X$, and we also extend $R_{ijkl}$ to the whole product~$X^{\times 10}$ by assuming that it acts identically on the six remaining spaces.

The \emph{set-theoretical four-simplex equation} (set-theoretical FSE) is, by definition, the following equality between two (compositions of) maps $X^{\times 10}\to X^{\times 10}$:
\begin{equation}\label{stfse}
R_{1234}R_{1567}R_{2589}R_{3680}R_{4790} = R_{4790}R_{3680}R_{2589}R_{1567}R_{1234}
\end{equation}
(here zero stands, of course, for 10).

\subsection{Quantum FSE}\label{ss:qfse}

Another version of FSE is \emph{quantum} equation. This can be written symbolically as the same equation~\eqref{stfse}, but with a different meaning of the symbols. Namely, let now $R$ be a linear operator acting in the fourth \emph{tensor} degree~$V^{\otimes 4}$ of a linear space~$V$. The quantum FSE is an equality between two operators in~$V^{\otimes 10}$. Similarly to Subsection~\ref{ss:stfse}, we call $V_1,\dots,V_{10}$ the ten copies of~$V$; $R_{ijkl}$ is the copy of~$R$ acting in $V_i\otimes V_j\otimes V_k\otimes V_l$ and also in the whole~$V^{\otimes 10}$, extended there by multiplying tensorially by identity operators in the remaining spaces.

\subsection{Nonconstant quantum FSE}\label{ss:ncfse}

Now we must confess that \eqref{stfse} is far from being the most general form of FSE. Below we call such equations (either set-theo\-retical or quantum) \emph{constant}, because each of them contains five copies of one and the same non-changing~$R$. While constant \emph{set-theoretical} FSE will be enough for us in this paper, we will be building from it solutions to a more complicated quantum equation, where there are five \emph{different} linear operators --- acting, however, in the same tensor products $V_i\otimes V_j\otimes V_k\otimes V_l$ as before. We will like also to change notations to \emph{calligraphic} letters for the tensor case, in order not to confuse it with the set-theo\-ret\-i\-cal case. So, the \emph{nonconstant} quantum FSE reads, by definition, as follows:
\begin{equation}\label{ncq}
\mathcal R_{1234}\mathcal S_{1567}\mathcal T_{2589}\mathcal U_{3680}\mathcal V_{4790} = \mathcal V_{4790}\mathcal U_{3680}\mathcal T_{2589}\mathcal S_{1567}\mathcal R_{1234}.
\end{equation}

\section{From set-theoretical to quantum FSE}\label{s:h}

\subsection{Permutation-type solutions}\label{ss:perm}

Let now there be a bijection $x\mapsto e_x$ between the set $X\ni x$ and a \emph{basis} of linear space~$V$. If $R$ is a solution to set-theoretic FSE~\eqref{stfse}, then we can get, in an obvious way, a solution~$\mathcal R$ to the constant quantum FSE (where $\mathcal V=\mathcal U=\mathcal T=\mathcal S=\mathcal R$), setting, by definition,
\[
\mathcal R(e_x\otimes e_y\otimes e_z\otimes e_t) = e_{x'}\otimes e_{y'}\otimes e_{z'}\otimes e_{t'},
\]
where
\[
(x',y',z',t') = R(x,y,z,t).
\]
Hietarinta~\cite{hietarinta} calls this \emph{permutation-type} solutions to (constant) quantum FSE.

\subsection{Cohomologies}\label{ss:coh}

Permutation-type solutions to constant quantum FSE can be generalized as follows. Set
\begin{equation}\label{ch}
\mathcal R(e_x\otimes e_y\otimes e_z\otimes e_t) = \varphi(x,y,z,t)e_{x'}\otimes e_{y'}\otimes e_{z'}\otimes e_{t'},
\end{equation}
where $\varphi$ is some scalar function. Operator~\eqref{ch} will satisfy the constant quantum equation provided function~$\varphi$ satisfies a system of as many as $(\# X)^{10}$ equations (that is, 1024 if $X$ contains two elements), see~\cite{KST}. We do not write them out here; what is important is that they are all of the multiplicative form
\begin{equation}\label{mc}
\varphi(\ldots)\varphi(\ldots)\varphi(\ldots)\varphi(\ldots)\varphi(\ldots) = \varphi(\ldots)\varphi(\ldots)\varphi(\ldots)\varphi(\ldots)\varphi(\ldots),
\end{equation}
where dots stay for some quadruples of arguments depending on~$R$.

Functions~$\varphi$ are called \emph{cochains} in~\cite{KST}. If it satisfies~\eqref{mc}, it is called \emph{cocycle}. Some of cocycles are, however, of little interest.

First, such are constant functions (taking just one fixed value). So, it makes sense to consider cocycles taken up to a constant (nonzero) factor, call them \emph{reduced} cocycles.

Second, such are \emph{coboundaries}, that is, functions of the form
\[
\varphi(x,y,z,t)=\frac{\psi(x')}{\psi(x)} \frac{\psi(y')}{\psi(y)} \frac{\psi(z')}{\psi(z)} \frac{\psi(t')}{\psi(t)}.
\]

So, of interest can be reduced cocycles modulo coboundaries, they may be called \emph{reduced homologies}. It seems, however, that nontrivial reduced homologies seldom exist in this setting.

\subsection{``Nonconstant cohomologies''}\label{ss:ncc}

The situation changes if we allow our quantum $\mathcal R$-operators to be different, as in equation~\eqref{ncq}. Let there be \emph{five different} functions $\varphi_{\mathcal R},\dots,\varphi_{\mathcal V}$, and set, instead of~\eqref{ch},
\begin{equation}\label{nch}
 \begin{array}{c}
\mathcal R(e_x\otimes e_y\otimes e_z\otimes e_t) = \varphi_{\mathcal R}(x,y,z,t)e_{x'}\otimes e_{y'}\otimes e_{z'}\otimes e_{t'},\\
\quad \dotfill \quad \\
\mathcal V(e_x\otimes e_y\otimes e_z\otimes e_t) = \varphi_{\mathcal V}(x,y,z,t)e_{x'}\otimes e_{y'}\otimes e_{z'}\otimes e_{t'}.
 \end{array}
\end{equation}

We call a quintuple $\varphi_{\mathcal R},\dots,\varphi_{\mathcal V}$ \emph{cocycle} if \eqref{ncq} holds; this happens provided it satisfies a system of $(\# X)^{10}$ equations of the form
\begin{multline}\label{phi}
\varphi_{\mathcal R}(\ldots)\varphi_{\mathcal S}(\ldots)\varphi_{\mathcal T}(\ldots)\varphi_{\mathcal U}(\ldots)\varphi_{\mathcal V}(\ldots) \\
= \varphi_{\mathcal V}(\ldots)\varphi_{\mathcal U}(\ldots)\varphi_{\mathcal T}(\ldots)\varphi_{\mathcal S}(\ldots)\varphi_{\mathcal R}(\ldots).
\end{multline}
There are, again, two kinds of trivial cocycles. First kind appears when each of $\varphi_{\mathcal R},\dots,\varphi_{\mathcal V}$ in constant (but, in contrast to Subsection~\ref{ss:coh}, there can be now five different constants). Second kind appears from \emph{ten} scalar functions on~$X$, call them $\psi_1,\dots,\psi_{10}$, and reads
\begin{equation}\label{iR}
\varphi_{\mathcal R}(x,y,z,t)=\frac{\psi_1(x')}{\psi_1(x)} \frac{\psi_2(y')}{\psi_2(y)} \frac{\psi_3(z')}{\psi_3(z)} \frac{\psi_4(t')}{\psi_4(t)},
\end{equation}
\begin{equation}\label{iS}
\varphi_{\mathcal S}(x,y,z,t)=\frac{\psi_1(x')}{\psi_1(x)} \frac{\psi_5(y')}{\psi_5(y)} \frac{\psi_6(z')}{\psi_6(z)} \frac{\psi_7(t')}{\psi_7(t)},
\end{equation}
\begin{equation}\label{iT}
\varphi_{\mathcal T}(x,y,z,t)=\frac{\psi_2(x')}{\psi_2(x)} \frac{\psi_5(y')}{\psi_5(y)} \frac{\psi_8(z')}{\psi_8(z)} \frac{\psi_9(t')}{\psi_9(t)},
\end{equation}
\begin{equation}\label{iU}
\varphi_{\mathcal U}(x,y,z,t)=\frac{\psi_3(x')}{\psi_3(x)} \frac{\psi_6(y')}{\psi_6(y)} \frac{\psi_8(z')}{\psi_8(z)} \frac{\psi_{10}(t')}{\psi_{10}(t)},
\end{equation}
\begin{equation}\label{iV}
\varphi_{\mathcal V}(x,y,z,t)=\frac{\psi_4(x')}{\psi_4(x)} \frac{\psi_7(y')}{\psi_7(y)} \frac{\psi_9(z')}{\psi_9(z)} \frac{\psi_{10}(t')}{\psi_{10}(t)}.
\end{equation}

\section{Calculations for Hietarinta's solutions}

\subsection[Two-color $\mathbb F_2$-linear solutions]{Two-color $\boldsymbol{\mathbb F_2}$-linear solutions}\label{ss:h}

Let now $X$ be the field $X=\mathbb F_2$ of two elements, that is, as a set, $X=\{0,1\}$. Assume also that $R$ is \emph{$\mathbb F_2$-linear}. In this situation, Hietarinta~\cite[Subsection~6.21]{hietarinta} calculated in 1997 all maps~$R$ enjoying~\eqref{stfse}. These linear maps are given by the following matrices:
\begin{align}
A_1=\begin{pmatrix} 0 & 1 & 0 & 1 \\ 1 & 0 & 1 & 0 \\ 0 & 0 & 0 & 1 \\ 0 & 0 & 1 & 0 \end{pmatrix},\qquad
A_2=\begin{pmatrix} 0 & 1 & 1 & 1 \\ 1 & 0 & 1 & 1 \\ 0 & 0 & 1 & 0 \\ 0 & 0 & 0 & 1 \end{pmatrix}, \label{A1,A2} \\[1ex]
A_3=\begin{pmatrix} 1 & 1 & 1 & 0 \\ 0 & 0 & 1 & 0 \\ 0 & 1 & 0 & 0 \\ 0 & 1 & 1 & 1 \end{pmatrix},\qquad
A_4=\begin{pmatrix} 1 & 1 & 1 & 1 \\ 0 & 0 & 1 & 1 \\ 0 & 1 & 0 & 1 \\ 0 & 0 & 0 & 1 \end{pmatrix},
\end{align}
and their transposes $A_1^{\mathrm T}$, $A_2^{\mathrm T}$, $A_3^{\mathrm T}$ and $A_4^{\mathrm T}$.

\subsection{Linear systems for logarithms}\label{ss:log}

In terms of logarithms $\log\varphi_{\ldots}$, all equations~\eqref{phi} become \emph{linear}. So do also \eqref{iR}--\eqref{iV}, in terms of $\log\varphi_{\ldots}$ and~$\log\psi_{\ldots}$. To specify each of five functions~$\varphi_{\ldots}$, we must specify its 16 values for 16 possible quadruples of arguments. So, equations~\eqref{phi} make a system of 1024 linear (in terms of logarithms) equations on 80 variables. Nontrivial solutions appear after factorizing modulo five additive constants for all $\log\varphi_{\ldots}$, and modulo those~$\varphi_{\ldots}$ that can be obtained in the form \eqref{iR}--\eqref{iV}.

\subsection{Computer calculations}\label{ss:calc}

The following two dimensions of linear spaces have been calculated using computer algebra system Maxima\footnote{http://maxima.sourceforge.net/ }:
\begin{itemize}
 \item the number~$n$ of independent equations among the 1024 equations~\eqref{phi},
 \item the dimension~$d$ of the linear space of quintuples $\log\varphi_{\mathcal R},\dots,\log\varphi_{\mathcal V}$ that can be obtained in the form \eqref{iR}--\eqref{iV}.
\end{itemize}
Then, the space of our ``nonconstant homologies'', implying nontrivial solutions to system of 1024 equations~\eqref{phi}, has dimension
\[
h=80-n-d-5,
\]
where $5$ stands for five additive constants mentioned above in Subsection~\ref{ss:log}.

The results are as follows.

\subsubsection*{For matrix $A_1$:}
\[
n=50, \qquad\qquad d=9, \qquad\qquad h=16.
\]

\subsubsection*{For matrix $A_2$:}
\[
n=37, \qquad\qquad d=7, \qquad\qquad h=31.
\]

\subsubsection*{For matrix $A_3$:}
\[
n=54, \qquad\qquad d=9, \qquad\qquad h=12. 
\]

\subsubsection*{For matrix $A_4$:}
\[
n=50, \qquad\qquad d=9, \qquad\qquad h=16.
\]

\subsubsection*{For matrix $A_1^{\mathrm T}$:}
\[
n=50, \qquad\qquad d=9, \qquad\qquad h=16.
\]

\subsubsection*{For matrix $A_2^{\mathrm T}$:}
\[
n=50, \qquad\qquad d=9, \qquad\qquad h=16.
\]

\subsubsection*{For matrix $A_3^{\mathrm T}$:}
\[
n=40, \qquad\qquad d=7, \qquad\qquad h=28.
\]

\subsubsection*{For matrix $A_4^{\mathrm T}$:}
\[
n=50, \qquad\qquad d=9, \qquad\qquad h=16.
\]

\section{From quantum FSE to quantum tetrahedron equation}\label{s:tetra}

Our results in Subsection~\ref{ss:calc} show that there exist large families of interesting solutions to quantum four-simplex equation~\eqref{ncq}. Now we can make from them solutions to the lower-dimensional quantum $n$-simplex equation, namely Zamolodchikov \emph{tetrahedron} (3-simplex) equation, by taking partial traces of our $\mathcal R$-operators in a well-known way. Even more interesting things come out if we apply more ingenuity and modify the traces by multiplying $\mathcal R$-operators by some additional operators, as we are going to explain. We will content ourself with doing our construction, in this paper, only for the matrix~$A_1$, see~\eqref{A1,A2}.

As all our vector spaces always have fixed bases, we do not make difference between operators and their matrices.

\subsection{Special cocycles leaving the last operator pure permutation}\label{ss:cs}

We construct such cocycles $(\varphi_{\mathcal R},\varphi_{\mathcal S},\varphi_{\mathcal T},\varphi_{\mathcal U},\varphi_{\mathcal V})$ for the permutation-type $\mathcal R$-operators generated by matrix~$A_1$ that $\varphi_{\mathcal V} \equiv 1$, that is, in terms of logarithms, find such subspace of the corresponding linear space where $\log\varphi_{\mathcal V} \equiv 0$. Our calculation using Maxima shows that its dimension is~14, while the dimension of the coboundary logarithms in the sense \eqref{iR}--\eqref{iU} is~6. There are also 4 constant cocycles, so, there remain $14-6-4=4$ essential parameters. 

Operator~$\mathcal V=\mathcal V_{4790}$ remains thus pure permutation, which implies the following important fact: \emph{$\mathcal V$ commutes with the operator}
\begin{equation}\label{1111}
\mathcal P = \mathcal P_{4790} = \begin{pmatrix} 1 & 1 \\ 1 & 1 \end{pmatrix}_4 \otimes \begin{pmatrix} 1 & 1 \\ 1 & 1 \end{pmatrix}_7 \otimes \begin{pmatrix} 1 & 1 \\ 1 & 1 \end{pmatrix}_9 \otimes \begin{pmatrix} 1 & 1 \\ 1 & 1 \end{pmatrix}_0. 
\end{equation}
This is because $\mathcal P$ can be also represented as a tensor product of a column matrix consisting of unities and a row matrix consisting of unities, and any permutation does not change a vector of unities. The subscripts in~\eqref{1111} mean of course the numbers of spaces, and tensor multiplication is implied by the identity matrices/operators $\begin{pmatrix} 1 & 0 \\ 0 & 1 \end{pmatrix}$ in the remaining six spaces.

Also, our calculations show that, for any individual operator $\mathcal R$, $\mathcal S$, $\mathcal T$ or~$\mathcal U$, cocycles give a 6-dimensional linear space of parameters, and coboundaries give a 3-dimensional space. As there is also one trivial multiplicative constant, there remain two essential parameters for an individual $\mathcal R$-operator.

\subsection{Tetrahedron solutions from special cocycles}\label{ss:t}

It follows from \eqref{ncq} and the commutativity between $\mathcal V$ and~$\mathcal P$ that
\begin{equation}\label{ncq'}
\mathcal P_{4790}\mathcal R_{1234}\mathcal S_{1567}\mathcal T_{2589}\mathcal U_{3680} = \mathcal V_{4790}\mathcal P_{4790}\mathcal U_{3680}\mathcal T_{2589}\mathcal S_{1567}\mathcal R_{1234}\mathcal V_{4790}^{-1}.
\end{equation}
Taking the partial trace of~\eqref{ncq'} in spaces 4, 7, 9 and~0, we arrive at the tetrahedron equation (with a somewhat nonstandard numbering of spaces)
\begin{equation}\label{KLMN}
\mathcal K_{123}\mathcal L_{156}\mathcal M_{258}\mathcal N_{368} = \mathcal N_{368}\mathcal M_{258}\mathcal L_{156}\mathcal K_{123}
\end{equation}
for operators
\begin{multline*}
\mathcal K_{123}=\Trace_4 \bigl( \begin{pmatrix} 1 & 1 \\ 1 & 1 \end{pmatrix}_4 \mathcal R_{1234} \bigr), \qquad
\mathcal L_{156}=\Trace_7 \bigl( \begin{pmatrix} 1 & 1 \\ 1 & 1 \end{pmatrix}_7 \mathcal S_{1567} \bigr), \\
\mathcal M_{258}=\Trace_9 \bigl( \begin{pmatrix} 1 & 1 \\ 1 & 1 \end{pmatrix}_9 \mathcal T_{2589} \bigr), \qquad
\mathcal N_{368}=\Trace_0 \bigl( \begin{pmatrix} 1 & 1 \\ 1 & 1 \end{pmatrix}_0 \mathcal U_{3680} \bigr).
\end{multline*}
Here $\Trace_k$ means of course the partial trace in the $k$-th space.

\subsection[Explicit form of tetrahedral $\mathcal R$-operators]{Explicit form of tetrahedral $\boldsymbol{\mathcal R}$-operators}\label{ss:R}

Direct calculations using Maxima lead to the following remarkable results. Fist, the number of essential parameters in our solution of equation~\eqref{KLMN} remains equal to~4 if we impose on matrices $\mathcal K,\mathcal L,\mathcal M,\mathcal N$ the additional requirement of being symmetric: $\mathcal K=\mathcal K^{\mathrm T}$, $\ldots$\,, $\mathcal N=\mathcal N^{\mathrm T}$. With this condition, it turns out that
\begin{itemize}
 \item $\mathcal K$ and~$\mathcal L$ are \emph{proportional}, or simply can be taken equal, and have the following form:
\begin{equation}\label{KL}
\mathcal K=\mathcal L=\begin{pmatrix} a&0&0&0&0&b&0 &0\\
 0&0&a&0&0&0&0&b \\
 0&a&0&0&b&0&0&0\\ 
 0&0&0&a&0&0&b&0\\
 0&0& b&0&0&0&0&c\\
 b&0&0&0&0&c&0&0\\
 0&0&0&b&0&0&c&0\\
 0&b&0&0&c&0&0&0 \end{pmatrix},
\end{equation}
 \item the same applies to matrices $\mathcal M$ and~$\mathcal N$:
\begin{equation}\label{MN}
\mathcal M=\mathcal N=\begin{pmatrix} a'&0&0&0&0&b'&0&0\\
 0&0&b'&0&0&0&0&c'\\
 0&b'&0&0&a'&0&0&0\\
 0&0&0&c'&0&0&b'&0\\
 0&0&a'&0&0&0&0&b'\\
 b'&0&0&0&0&c'&0&0\\
 0&0&0&b'&0&0&a'&0\\
 0&c'&0&0&b'&0&0&0 \end{pmatrix},
\end{equation}
 \item equality~\eqref{KLMN} imposes \emph{no} dependencies on parameters $a,b,c,a',b',c'$. So, six of them minus two scalings in \eqref{KL} and~\eqref{MN} give four essential parameters as promised.
\end{itemize}

\section{Algebraic nontriviality, but simple thermodynamical behavior}\label{s:a}

\subsection{Genuine three-dimensionness}\label{ss:gtd}

Let us speak, for concreteness, of operator~$\mathcal K$. It acts in the tensor product $V_1\otimes V_2\otimes V_3$ of three two-dimensional linear spaces with fixed bases. Here we show  that there is no such vector
\begin{equation}\label{u}
\begin{pmatrix}1\\ u\end{pmatrix}\in V_1
\end{equation}
that
\begin{equation}\label{uX}
\begin{pmatrix}1\\ u\end{pmatrix} \otimes X \stackrel{\mathcal K}{\mapsto} \begin{pmatrix}1\\ u\end{pmatrix} \otimes Y
\end{equation}
for any vector~$X\in V_2\otimes V_3$ in the tensor product of two remaining spaces. Of course, similar facts can be also proved for $\begin{pmatrix}1\\ v\end{pmatrix}\in V_2$ and $\begin{pmatrix}1\\ w\end{pmatrix}\in V_3$.

If \eqref{uX} held true, then $\mathcal K$ could be thought of as not completely three-dimen\-sional (in the sense ``well suited for three-dimensional lattice integrable models in mathematical physics''), because at least if vector~\eqref{u} is on its first input, then $\mathcal K$ would reduce to the two-dimensional operator $K_{23}\colon\;X\mapsto Y$. More formally, $\mathcal K$ would acquire block structure $\begin{pmatrix}A&B\\ 0&C\end{pmatrix}$ if we change the basis in~$V_1$ properly; $A$, $B$ and~$C$ are here $4\times 4$ blocks. This block structure would, in its turn, imply that some nontrivial linear combination of $4\times 4$ blocks in~\eqref{KL} is zero matrix --- but this is obviously impossible.

\subsection{No disguised permutation}\label{ss:ndp}

Our tetrahedral $\mathcal R$-operators do not turn into a permutation with possible cocycle multipliers (tetrahedral analogue of~\eqref{ch}; below ``permutation with cocycle'' for short) under any ``gauge'' transformation. That is, if we speak again of operator~$\mathcal K$, let $F,G,H$ be any invertible $2\times 2$ matrices, then
\[
(F\otimes G\otimes H)\, \mathcal K\, (F\otimes G\otimes H)^{-1}
\]
is not a permutation with cocycle.

If $\mathcal K$ \emph{were} such a permutation with cocycle, this could be reformulated as follows: there are vectors
\begin{equation}\label{v}
f_i=\begin{pmatrix}1\\ u_i\end{pmatrix}\in V_1,\qquad g_j=\begin{pmatrix}1\\ v_j\end{pmatrix}\in V_2,\qquad h_k=\begin{pmatrix}1\\ w_k\end{pmatrix}\in V_3,\qquad i,j,k=1,2,
\end{equation}
forming bases in their corresponding spaces and such that
\begin{equation}\label{uvw}
\mathcal K(u_i\otimes v_j\otimes w_k) = \varphi_{ijk} (u_{i'}\otimes v_{j'}\otimes w_{k'})
\end{equation}
for any $i,j,k$ --- that is, vectors~\eqref{v} would be ``vacuum vectors'' of~$\mathcal K$ in the Krichever's~\cite{krichever} sense.

Some degree of any permutation makes identical mapping. Considering the 1st and 6th components of vectors in~\eqref{uvw} in one case and the 4th and 7th components in the other, and taking into account the explicit form~\eqref{KL} of~$\mathcal K$, we see that $\begin{pmatrix}1\\ u_iw_j\end{pmatrix}$ and $\begin{pmatrix}w_k\\ u_l\end{pmatrix}$ are \emph{eigenvectors for some degree of matrix~$\begin{pmatrix}a&b\\ b&c\end{pmatrix}$} and hence (as $a,b,c$ are free parameters) for this matrix itself, for all $i,j,k,l$. Simple analysis, using the fact that the mentioned matrix has not more than two eigenvectors, shows that this cannot be.

\subsection{Thermodynamical behavior}\label{ss:thb}

Two obvious eigenvectors of operator~\eqref{KL} are
\[
\Omega_{1,2} = \begin{pmatrix}1\\ u_{1,2}\end{pmatrix}\otimes \begin{pmatrix}1\\ 1\end{pmatrix}\otimes \begin{pmatrix}1\\ 1\end{pmatrix},
\]
where $\begin{pmatrix}1\\ u_{1,2}\end{pmatrix}$ are two eigenvectors of matrix~$\begin{pmatrix}a&b\\ b&c\end{pmatrix}$. Let now $a$, $b$ and~$c$ in~\eqref{KL} be all \emph{positive}, then exactly one of~$u_{1,2}$ is also positive, let it be~$u_1$, and call the corresponding eigenvalue~$\lambda$.

Let there be a statistical physical model on a cubic lattice, with operator~$\mathcal K$ in each vertex. Cutting the lattice into slices by planes orthogonal to a space diagonal, we obtain a ``hedgehog'' transfer matrix in each slice, see~\cite{PS}. The positive eigenvector of such transfer matrix --- the only one essential for the thermodynamical limit --- is a tensor product of vectors~$\Omega_1$, and this leads to the simple fact that the free energy per site (vertex) in the thermodynamical limit is~$\log \lambda$.

\section{Discussion}

Our general idea for searching of solutions to simplex equations can now be stated as follows. Take a \emph{higher} simplex equation, find a Hietarinta-style ``linear permutation'' solution for it, then calculate its ``nonconstant homologies'', and then descend to the required simplex dimension by taking partial traces and possibly doing additional tricks, like in our Section~\ref{s:tetra}. We did it for 3-simplex equation and using 4-simplex equation, and this already looks like a useful step in the right direction: the obtained solutions of tetrahedron equation, as explained in Section~\ref{s:a}, look significantly richer than simple permutations. Our hope is that even more impressive results could be obtained if we employ simplex equations of great dimensions, for instance, $n=100500$ or even $n=\infty$, as Hietarinta already proposed in~\cite[Section~7]{hietarinta}. Perhaps, interesting statistical physical models, with nonnegative Boltzmann weights and nontrivial thermodynamical behavior, are waiting for us on this way.

\end{document}